\documentstyle[preprint,aps,graphicx]{revtex}
\tightenlines
\begin{document}
\newcommand{\bref}[1]{eq.~(\ref{#1})}
\newcommand{\be}{\begin{equation}}
\newcommand{\en}{\end{equation}}
\newcommand{\bs}{$\backslash$}
\newcommand{\us}{$\_$}

\title{Theoretical model for magnetooptic imaging}
\author{L.E. Helseth}
\address{Department of Physics, University of Oslo, P.O. Box 1048 Blindern,
N-0316 Oslo, Norway\\
email: l.e.helseth@fys.uio.no
}

\maketitle
\begin{abstract}
A theoretical model for magnetooptic imaging is presented. The model gives a
detailed description of the magnetooptic indicator and the optical imaging 
system, where the intensity at the detector plane is described by a set of 
diffraction integrals. To demonstrate the applicability of the model, some 
issues concerning imaging of single vortices in superconductors are studied. 
It is found that the useful signal is strongly dependent on the penetration 
depth of the superconductor, and also the sensitivity and thickness of the 
indicator. 

\end{abstract}

\narrowtext

\newpage

\section{Introduction}
For centuries humans have tried to 'see' magnetic fields using various devices.
Iron particles, magnetic fluids, inductive coils, Hall probes and magnetic 
bacterias are only a few of the approaches which have been adopted. During the 
past 40 years the research on magnetooptic visualization has been highly motivated by the need 
to 'see' magnetic fields from superconductors. To that end, magnetooptic
visualization using thin film indicators has been particularly useful. 
In 1957 P.B. Alers visualized for the first time the magnetic field 
distribution in a superconductor using a mixture of cerous 
nitrate-glycerol\cite{Alers}. A highly successful development 
followed in 1968 when H. Kirchner noticed that chalcogenide-films could 
be used to visualize fields with improved sensitivity\cite{Kirchner}. 
In the following 20 years magneto-optic studies were published periodically, 
mainly looking at flux profiles in superconductors\cite{Habermeier1,Habermeier2}. 
However, there were no major advances until 1991, when it was discovered that bismuth-substituted 
iron garnet films with in-plane magnetization are excellent indicators with almost no domain activity\cite{Antonov,Jevenko,Belyaeva}.
This discovery triggered a large number of quantitative investigations of the 
flux behaviour in high-temperature superconductors\cite{Dorosinskii,Koblischka,Polyanskii}. Two of the most recent 
advances in this field are ultrafast magnetooptical studies (see e.g
ref.\cite{Runge}) and imaging of single vortices\cite{Goa}.

Although magnetooptical indicators have mainly been used to image magnetic
fields from superconductors, a large number of other possible application areas
exists. Thus, it has been realised that such indicators are useful for 
imaging magnetic defects, magnetic phase transitions, microelectronic circuits 
and magnetic storage media\cite{Helseth,Vlasko-Vlasov,Egorov,Grechishkin,Kotov,Shamonin}. 

Since the field of magnetooptic visualization has matured for more than 
40 years, one may expect that most parts of the image formation process is
well understood. This is not the case. In fact, only recently a  
model for magnetooptic imaging has been published\cite{Shamonin}. However, this model is
mostly concerned with the magnetization redistribution in the indicator. 
The purpose of the current paper is to develop a more detailed model for the 
magnetooptic imaging system (see Fig. \ref{f1}), including the indicator, and to use this
to examine the possibility of imaging weak magnetic fields, e.g. single vortices in 
superconductors. 
   
\section{Magnetization in the indicator}
The basic sensor element for detection of magnetic fields is the magnetooptic
indicator. It is therefore first necessary to know how the indicator respond 
to an external field\cite{Shamonin}.  
Let $\mbox{\boldmath $M$} $ be the magnetization vector in
the indicator, and $M_{s}$ be the saturation magnetization. Then we have from
Fig. \ref{f2}
\begin{equation}
\mbox{\boldmath $M$} =M_{s}[cos\beta cos\phi _{M}, cos\beta sin\phi _{M},
sin\beta]=[M_{r} cos\phi _{M}, M_{r}sin\phi _{M}, M_{z}]\,\,\, .
\end{equation}
The external field is given by
\begin{equation}
\mbox{\boldmath $H$} =[H_{r}cos\phi _{H} ,H_{r}sin\phi _{H}, H_{z}] \,\,\,  ,
\end{equation}
where $M_{r} =\sqrt{M_{x}^{2} +M_{y}^{2}}$ and $H_{r} =\sqrt{H_{x}^{2} +H_{y}^{2}}$.
We assume that the indicator consists of a single
domain, and that its free energy can be expressed as a sum of the 
uniaxial anisotropy energy ($E_{u}$), the Zeeman energy ($E_{z}$) and the 
demagnetizing energy ($E_{d}$)
\begin{equation}
E=E_{u} +E_{z} + E_{d} \,\,\,  .
\end{equation}
The uniaxial anisotropy energy can be written as,
\begin{equation}
E_{u}=K_{u}cos^{2}\beta \,\,\,  ,
\end{equation}
where $K_{u}$ is the uniaxial anisotropy constant. The Zeeman energy is given 
as
\begin{equation}
E_{z}= -\mu _{0}\mbox{\boldmath $M$} \mbox{\boldmath $\cdot$} \mbox{\boldmath
$H$} = -\mu _{0} M_{s}[H_{r}cos\beta cos(\phi _{M} -\phi _{H})  +H_{z}sin\beta ]
\,\,\,  .
\end{equation}
Since we neglect the cubic anisotropy of the indicator the 
in-plane magnetization direction must be the 
same as that of the external field ($\phi _{M} =\phi _{H}$). 
The demagnetizing energy, which tends to keep the magnetization in the plane of
the film, can be given as
\begin{equation}
E_{d}=\frac{\mu _{0} M_{s}^{2}}{2}sin^{2}\beta \,\,\,  .
\end{equation} 
 
Minimizing $E$ with respect to $\beta$ results in the following relationship
\begin{equation}
H_{a}sin\beta cos\beta + H_{r}sin\beta - H_{z}cos\beta=0 \,\,\, ,
\end{equation}
where the socalled anisotropy field is given by
\begin{equation}
H_{a}=M_{s} - \frac{2K_{u}^{tot}}{\mu _{0}M_{s}}\,\,\, .
\end{equation}
If we assume that $H_{r}=0$, then the resulting expression for the 
magnetization is
\begin{equation}
M_{r} = M_{s}\sqrt{ 1- \left( \frac{H_{z}}{H_{a}} \right) ^{2} } \,\,\,  ,H_{z}
\leq H_{a} \,\,\,  ,
\end{equation}
and
\begin{equation}
M_{z} = M_{s}\frac{H_{z}}{H_{a}} \,\,\,  , H_{z} \leq H_{a}\,\,\,  .
\end{equation}
If $\beta $ is very small (corresponding to very small $H_{z}$, or large $H_{a}$
or $H_{r}$), we may put $sin\beta \approx \beta$ and $cos\beta \approx 1$, and 
the magnetization becomes
\begin{equation}
M_{r} \approx M_{s} \,\,\,  , H_{z} \ll H_{a}\,\,\,  ,
\label{a}
\end{equation}
and
\begin{equation}
M_{z} \approx M_{s}\frac{H_{z}}{H_{a} +H_{r}} \,\,\,  , H_{z} \ll H_{a}\,\,\,  ,
\label{b}
\end{equation}

Thus we see that an in-plane component $H_{r}$ will only reduce the z-component
of the magetization. Therefore, it is necessary to keep $H_{r}$ and $H_{a}$
small if large sensitivity to $H_{z}$ is needed. 

\section{The magnetooptic imaging system}
The basic imaging system is shown in Fig.\ref{f1}. In principle, it is a
conventional polarizing microscope, although many different designs 
exists\cite{Goa}. Basically, the z-component (or any other component) of the 
magnetization is converted into a small rotation of the plane of polarization, 
which can then be detected by an array detector (e.g. CCD). The signal on the detector can be found by an 
analysis based on Jones matrices together with the methods used in 
refs.\cite{Novotny,Helseth1}. 
For a general polarization distribution incident on the objective we
write\cite{Helseth1}
\begin{displaymath}
\mathbf{E_{0}} = 
\left[ \begin{array}{ccc} 
a(\theta, \phi ) \\
b(\theta, \phi ) \\
\end{array} \right] \,\,\, ,
\end{displaymath}
where $\theta$ is the incident angle and $\phi $ is the azimuthal angle.  
In this paper we assume that the incident polarization after the 
polarizer consists of a x-component only (a=1 and b=0). The effect 
of the objective lens is to change the incident polarization into s and 
p-polarized light with the following transformation 
\begin{displaymath}
\mathbf{L} = 
\left[ \begin{array}{ccc} 
cos\phi & sin\phi \\
-sin\phi & cos\phi \\
\end{array} \right]\,\,\, .
\end{displaymath}
Next, the rays passes through the isotropic substrate on which the
indicator is deposited
\begin{displaymath}
\mathbf{T_{in}} = 
\left[ \begin{array}{ccc} 
t_{p} ^{in} & 0\\
0 & t_{s} ^{in} \\
\end{array} \right]\,\,\, .
\end{displaymath}
When the rays are reflected from the magnetooptic layer, some of the s-polarization is
transformed into p-polarization (and vice versa). In order to assure maximum 
reflection, one may evaporate a thin mirror onto the indicator. The 
reflection matrix from the indicator and the mirror can be expressed as
\begin{displaymath}
\mathbf{R} = 
\left[ \begin{array}{ccc} 
r_{pp} &r_{ps} \\
r_{sp} & r_{ss} \\
\end{array} \right]\,\,\, ,
\end{displaymath}
where $r_{pp}$ and $r_{ss}$ are the usual Fresnel coefficients.
In the first order approximation (corresponding to weak magnetooptic effects), 
we find the following off-axis components for arbitrary magnetization vectors 
and angles of incidence   
\begin{equation}
r_{ps} =\frac{\pi n_{1} Q}{\lambda _{0}} t_{01}^{s}t_{10}^{p}\int _{D}^{} f(z)
\left(\frac{a(z)M_{z} + b(z)M_{y} tan\theta _{1}}{M_{s}} \right) dz  \,\,\,  ,
\end{equation} 
\begin{equation}
r_{sp} =-\frac{\pi n_{1} Q}{\lambda _{0}}t_{01}^{s}t_{10}^{p}\int _{D}^{} f(z) 
\left(\frac{a(z)M_{z} - b(z)M_{y} tan\theta _{1}}{M_{s}} \right) dz \,\,\,  ,
\end{equation} 
where $n_{1}$ is the refractive index of indicator, Q is the magneto-optical material constant, $\lambda_{0}$ is the
wavelength in vacuum, $t_{ij}^{s,p}$ is the s or
p-component of the transmission coefficient through the interface from medium 
i to j (similar for $r_{ij}^{s,p}$), $\theta _{1}$ is the propagation angle in the
indicator, $f(z)=exp[-i4\pi n_{1}zcos\theta _{1} /\lambda _{0}]$, and 
\begin{equation}
a(z) = \frac{[1+r_{12}^{s}f(D-z)][1+r_{12}^{p}f(D-z)]}{[1-r_{10}^{s}r_{12}^{s}f(D)][1-r_{10}^{p}r_{12}^{p}f(D)]} \,\,\,  ,
\end{equation}
\begin{equation}
b(z) = \frac{[1+r_{12}^{s}f(D-z)][1-r_{12}^{p}f(D-z)]}{[1-r_{10}^{s}r_{12}^{s}f(D)][1-r_{10}^{p}r_{12}^{p}f(D)]}\,\,\,  ,
\end{equation}
according to the recipy of Hubert and Traeger\cite{Hubert}. The integration is
carried out over the thickness (D) of the indicator. Here we have
neglected the transverse Kerr-effect, since this is rather small in most
indicators. 

For strong magnetooptical effects, one should use a similar formalism, but take 
into account multiple scattering. This complicates the problem considerably, 
and will not be pursued here. 

On its way back to the array detector, the light will again pass through the
substrate
\begin{displaymath}
\mathbf{T_{out}} = 
\left[ \begin{array}{ccc} 
t_{p} ^{out} & 0\\
0 & t_{s} ^{out} \\
\end{array} \right] \,\,\,  .
\end{displaymath}
Furthermore, it is collected by the objective lens, and in total we can
write the electric field before the analyzer as
\begin{equation}
E=L^{-1}T_{out}RT_{in}LE_{0} \,\,\,  .
\end{equation}
After evaluating the equation above we observe that also the Fresnel 
coefficients between the substrate and air influences the polarization state 
of the outgoing beam. In many cases it is possible to deposit a dielectric coating which minimizes the
depolarization due to the substrate. Here we assume the most ideal case, in
which all transmission coefficients become unity. In this case we may write
\begin{equation}
E_{x} = \frac{1}{2} (r_{pp}+r_{ss}) +\frac{1}{2}(r_{pp}-r_{ss})cos2\phi -
\frac{1}{2}(r_{sp}+r_{ps})sin2\phi  \,\,\,  ,
\end{equation}
and
\begin{equation}
E_{y}=\frac{1}{2} (r_{sp}-r_{ps}) +\frac{1}{2} (r_{pp}-r_{ss}) sin2\phi +
\frac{1}{2}(r_{sp}+r_{ps})cos2\phi \,\,\,  .
\end{equation}
These equations can be used to calculate the expected polarization pattern
at the exit pupil seen in the conoscopic image, which is often used to judge the
quality of the imaging system. 
To find the electric field at the detector plane, ($r',\phi _{c}$), we assume
that the imaging system obeys the sine condition, and use 
the method of refs.\cite{Novotny,Helseth1} to obtain
\begin{equation}
E_{xD} = I_{0}^{a} +I_{2}^{a}cos2\phi _{c} + I_{2}^{b}sin2\phi _{c} \,\,\,  ,
\label{sig1}
\end{equation}
and
\begin{equation}
E_{yD} = I_{0}^{b} +I_{2}^{a}sin2\phi _{c} - I_{2}^{b}cos2\phi_{c}  \,\,\,  ,
\label{sig2}
\end{equation}
where
\begin{equation}
I_{0}^{a} = \int_{0}^{\alpha} (r_{pp}+r_{ss}) sin\theta cos\theta J_{0}
(kr'sin\theta ) exp(2ikzcos\theta ) d\theta \,\,\,  ,
\end{equation}

\begin{equation}
I_{0}^{b} = \int_{0}^{\alpha} (r_{sp}-r_{ps}) sin\theta cos\theta J_{0}
(kr'sin\theta ) exp(2ikzcos\theta ) d\theta  \,\,\,  ,
\end{equation}

\begin{equation}
I_{2}^{a} = \int_{0}^{\alpha} (r_{ss}-r_{pp}) sin\theta cos\theta J_{2}
(kr'sin\theta ) exp(2ikzcos\theta) d\theta \,\,\,  ,
\end{equation}

\begin{equation}
I_{2}^{b} = \int_{0}^{\alpha} (r_{sp}+r_{ps}) sin\theta cos\theta  J_{2}
(kr'sin\theta ) exp(2ikzcos\theta ) d\theta \,\,\,  . 
\end{equation}
Here $J_{n}$ is the Bessel function of the first kind (of order $n$), and  
a constant in front of the integrals is ommited. Note that the integral is
taken over the angular aperture of the objective lens, which has a numerical
aperture $NA=sin\alpha $. In deriving these formulas, we have assumed that the objective is corrected for
all spherical aberrations introduced by the film and the substrate.
Here the factor $cos\theta $ occurs since we have chosen aplanatic apodization
at the exit pupil ($\sqrt{cos\theta }$, squared for two passes through the
objective). The phase factor $exp(i2kzcos \theta )$ is included to account 
for the defocus of the light beam. Here we will assume that 
z=0, and therefore that the light beam is at focus on the mirror-plane.  
When an analyzer is placed in front of the detector, the intensity 
at the detector plane must be written as 
\begin{equation}
i_{D} = |E_{xD}cos\gamma + E_{yD}sin\gamma |^{2} \,\,\, ,
\label{inte}
\end{equation}
where $\gamma$ is the analyzer's angle from the x-axis.    
So far, we have not considered diffraction due to the localized magnetization
distribution in the indicator. However, this can be included by using the procedure of Kambersky
et al.\cite{Kambersky}. In that paper a procedure for calculating the 
diffracted Kerr or Faraday amplitude is presented. The result 
must be weighted by the polarization vector of the lens system, and summed up 
at the exit pupil. This is a large numerical task, 
at least for a reflection-type microscope, and will not be 
pursued here. That is, we will neglect diffraction from magnetization 
gradients in the indicator.

\section{Imaging of single vortices in a superconductor}
Only recently magnetooptic imaging of single vortices was achieved
experimentally\cite{Goa}. In this section we study magnetooptic imaging 
of single vortices using approximative solutions of the theory given above.
Interestingly, it has been found that the field from a
vortex is similar to that from a magnetic monopole located a distance 
$z_{0}=-1.27\lambda $ ($\lambda $ is the penetration depth) below the 
superconductor surface\cite{Carneiro}. In this approximation the scalar 
potential from a vortex can be written as
\begin{equation}
\phi _{V} =\frac{\Phi_{0}}{2\pi \mu _{0}} \frac{1}{\sqrt{x^{2} + y^{2}
+(z-z_{0})^{2}}} \,\,\,  ,
\label{a}
\end{equation}
where $\Phi _{0}$ is the flux quantum. 

To get some insight into the behaviour of the magnetooptic response, let 
us neglect the absorption and the multiple reflections, and 
assume that $H_{a} \gg H_{r},\,H_{z} $. Then the off-diagonal components of the 
reflection becomes 
\begin{equation}
r_{ps} \propto Q t_{01}^{s} t_{10}^{p} \left \{ \frac{1}{H_{a} } 
\left[ \phi _{V} (d) -\phi _{V} (d+D) \right] + Dtan\theta _{1}sin\phi _{M} \right \} \,\,\,  ,
\end{equation}  
and
\begin{equation}
r_{sp} \propto - Q t_{01}^{s} t_{10}^{p} \left \{
\frac{1}{H_{a} } \left[ \phi _{V} (d) -\phi _{V} (d+D) \right] - Dtan\theta _{1} sin\phi _{M} \right \} \,\,\,  ,
\end{equation}
where $d$ is the distance between the superconductor and the indicator.
Although these expressions are approximations, they still contain some
interesting physics. 
Note that the second term is constant, i.e. independent of the vortex field, 
since we have assumed that $H_{a} \gg H_{r},\,H_{z}$. 
It is also seen that the off-axis reflection coefficients are strongly dependent on the penetration 
depth of the superconductor. That is, when the penetration depth is large, 
the magnetic monopole appears to be located far from the surface of the 
superconductor, whereas when it is small it is located close to the surface. 
Thus the signal increases with decreasing penetration depth. As an example of 
this, Fig. \ref{f3} shows $r_{ps}$ (without the constant term) as a function of the penetration depth when 
d=125 nm (dashed line) and d=250 nm (solid line). It is noted that the 
magnetooptical signal falls off nonlinearly with the penetration depth, but 
that the gradient is less severe when the distance between the superconductor 
and the indicator increases.     

To find the actual signal at the array detector, one must evaluate 
Eq. \ref{inte}. When the polarizer and analyzer are in nearly 
crossed positions ($\alpha \approx 90 ^{\circ}$), and $\phi _{c} =0^{\circ}$, 
the signal can be approximated by 
\begin{equation}
i_{D} \propto Q^{2} \left[ \phi _{V}^{im} (d)  -\phi _{V} ^{im} (d+D) \right] ^{2} |\int_{0}^{\alpha}
t_{01}^{s} t_{10}^{p} sin\theta  cos\theta  J_{0}
(kr'sin\theta ) d\theta |^{2} \,\,\,  .
\label{int}
\end{equation} 
Here $[\phi _{V} ^{im} (d) -\phi _{V}  ^{im}(d+D)]^{2}$ may be interpreted as the
geometric image of a given position ($r, \phi$) at the indicator. According to 
the Huygens-Fresnel principle, each such point generates a spherical secondary
wavelet. Since these wavelets must pass through the apertures of the optical 
system, the 'perfect' geometric image is smeared out at the image plane. That
is, each point in the geometric image must be multiplied by a diffraction
integral and the point becomes a distribution at the detector.  
One should also note that in the present approximation the diffraction integral 
is modified by the off-axis reflection coefficient through the 
factor $t_{01}^{s} t_{10}^{p}$. In this paper we have not considered the response of 
the array detector, as this is assumed to be ideal. 

It is very important to note that the signal is proportional to the 
difference in magnetic potential at the two surfaces of the
indicator. For this reason the signal increases with thickness, but the 
signal-distribution also broadens significantly. This issue will be 
of particular importance when the vortices are placed close together, and the
flux-profiles overlap. Thus one may expect a thin indicator to provide better 
signal contrast than a thick one, in particular since the contrast is often 
more important than the strength of the signal. To illustrate this point, 
Fig. \ref{f4} shows $[\phi _{V} ^{im}(d) -\phi _{V} ^{im}(d+D)]^{2} $ at the
detector plane when the vortices are placed 2 $\mu m$ apart, and the thickness 
of the indicator is D=5 $\mu m$ (dashed line) and D=1 $\mu m$ (solid line). 
Although the signal from the thickest films is 1.7 times stronger than that of 
the thinnest film, we see that the contrast is better in the latter case.

Finally, it is important to point out that in 
deriving Eq. (\ref{int}) we have done several assumptions in order to obtain a 
simple expression which is readily interpreted. If a detailed
analysis is required, one should adopt the full formalism presented in this
paper. It is our hope that this formalism could be useful for future studies 
on magnetooptic visualization.

\section{Conclusion}
A theoretical model for investigating magnetooptic imaging of
weak magnetic fields has been presented. It takes into account the details of 
the indicator and the magnetooptic system.  To demonstrate the applicability 
of the model, we have tried to understand some issues concerning imaging of 
single vortices in superconductors. It is found that the useful signal is 
strongly dependent on the penetration depth of the superconductor, and also the sensitivity and 
thickness of the indicator.

\acknowledgements
The author is grateful to P.E. Goa, H. Hauglin, M. Baziljevich and 
T.H. Johansen for many interesting discussions on this topic. This work was 
supported by the Norwegian Research Council and Tandberg Data ASA.

\begin{figure}
\caption{The basic setup for magnetooptical imaging.  
\label{f1}}
\end{figure}

\begin{figure}
\caption{The magnetization vector in the indicator.  
\label{f2}}
\end{figure}

\begin{figure}
\caption{The off-axis reflection coefficient as a function of the penetration
depth when d=125 nm (dashed line) and d=250 nm (solid line). Here D=1 $\mu m$.   
\label{f3}}
\end{figure}

\begin{figure}
\caption{The normalized signal from three vortices placed 2 $\mu m$ apart. Here $\lambda=$
100 nm, D=1 $\mu m$ (solid line) and D=5 $\mu m$ (dashed line). We have 
neglected diffraction and assumed a magnification M=1.    
\label{f4}}
\end{figure}

\newpage
\centerline{\includegraphics[width=14cm]{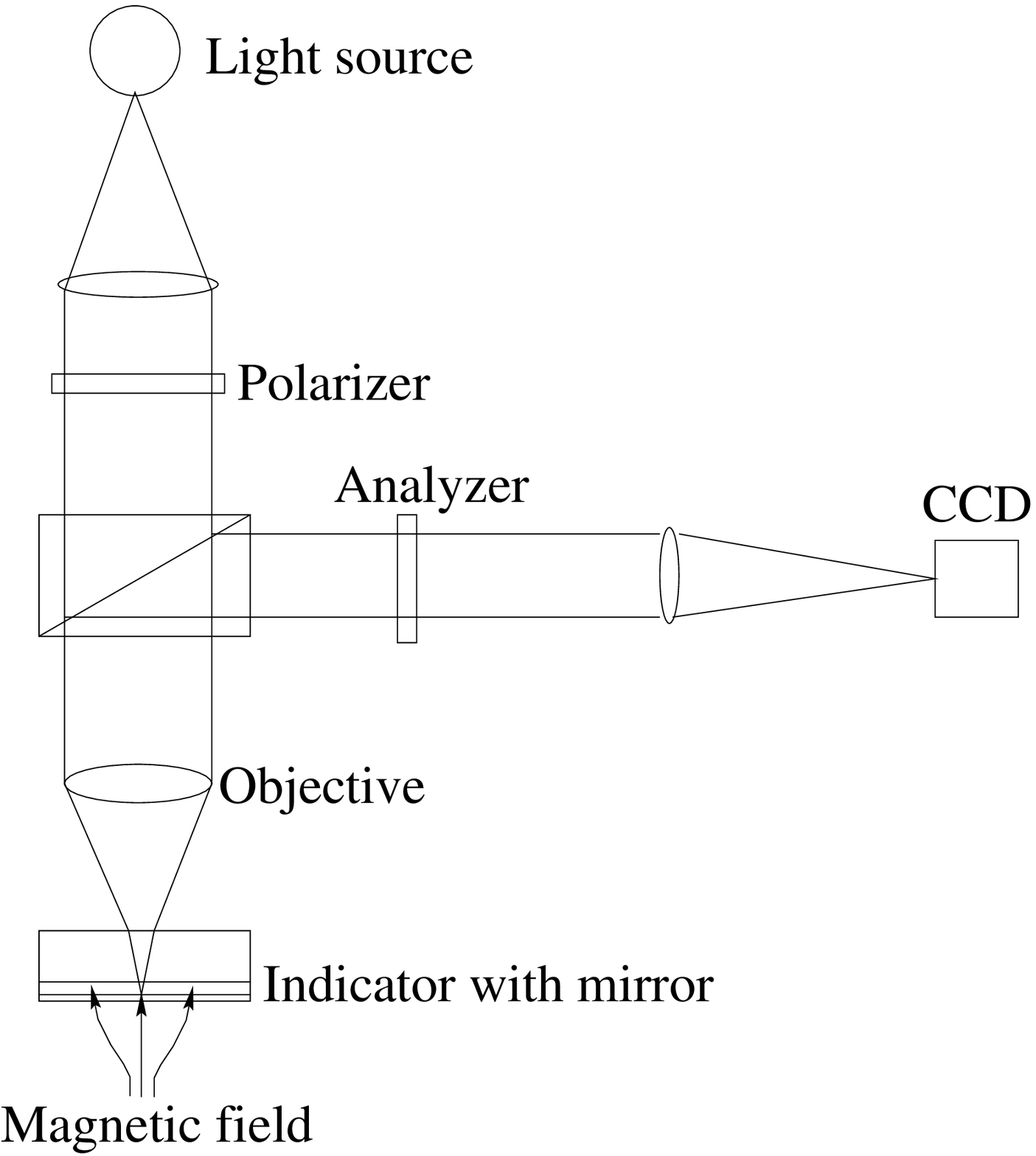}}
\vspace{2cm}
\centerline{Figure~\ref{f1}}

\newpage
\centerline{\includegraphics[width=14cm]{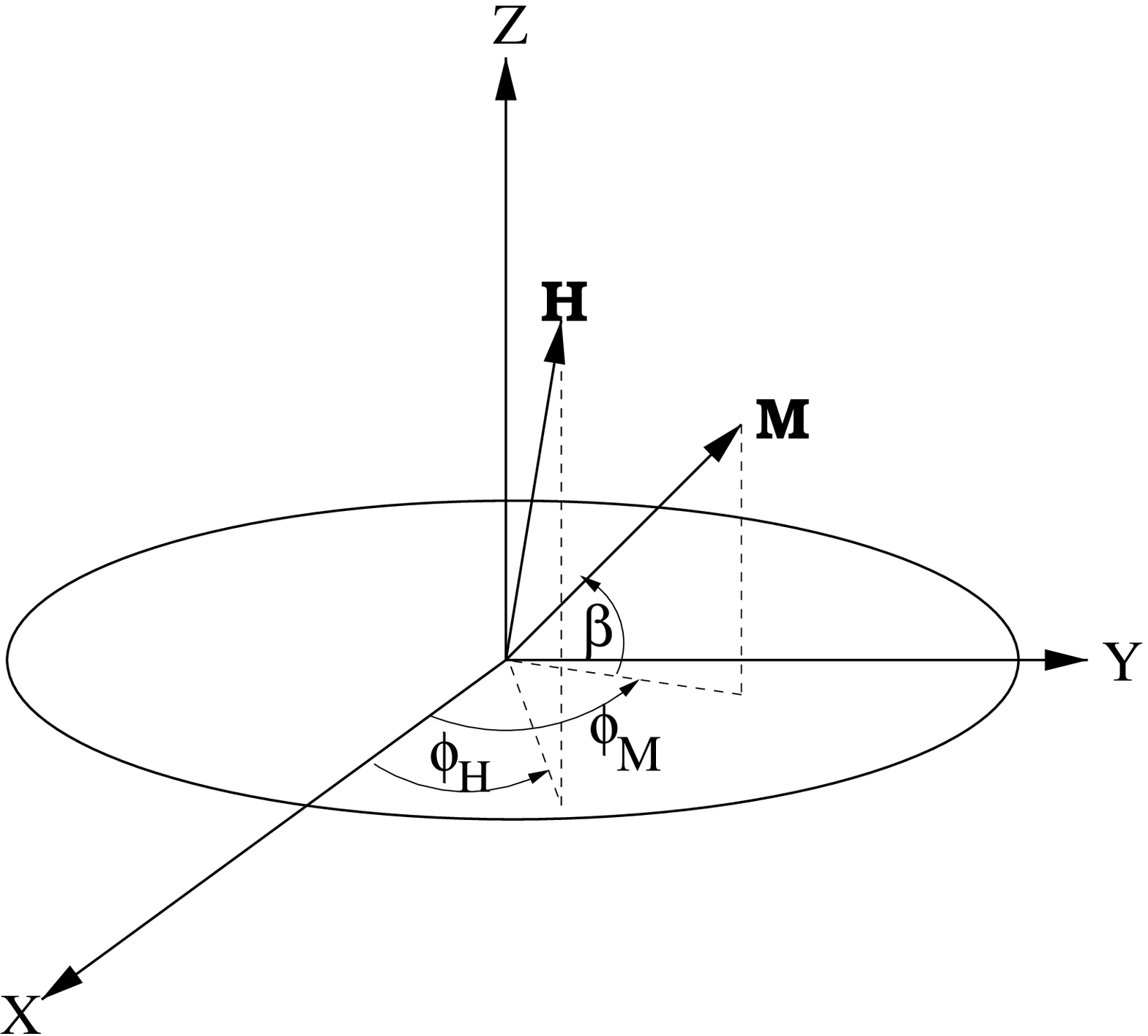}}
\vspace{2cm}
\centerline{Figure~\ref{f2}}

\newpage
\centerline{\includegraphics[width=14cm]{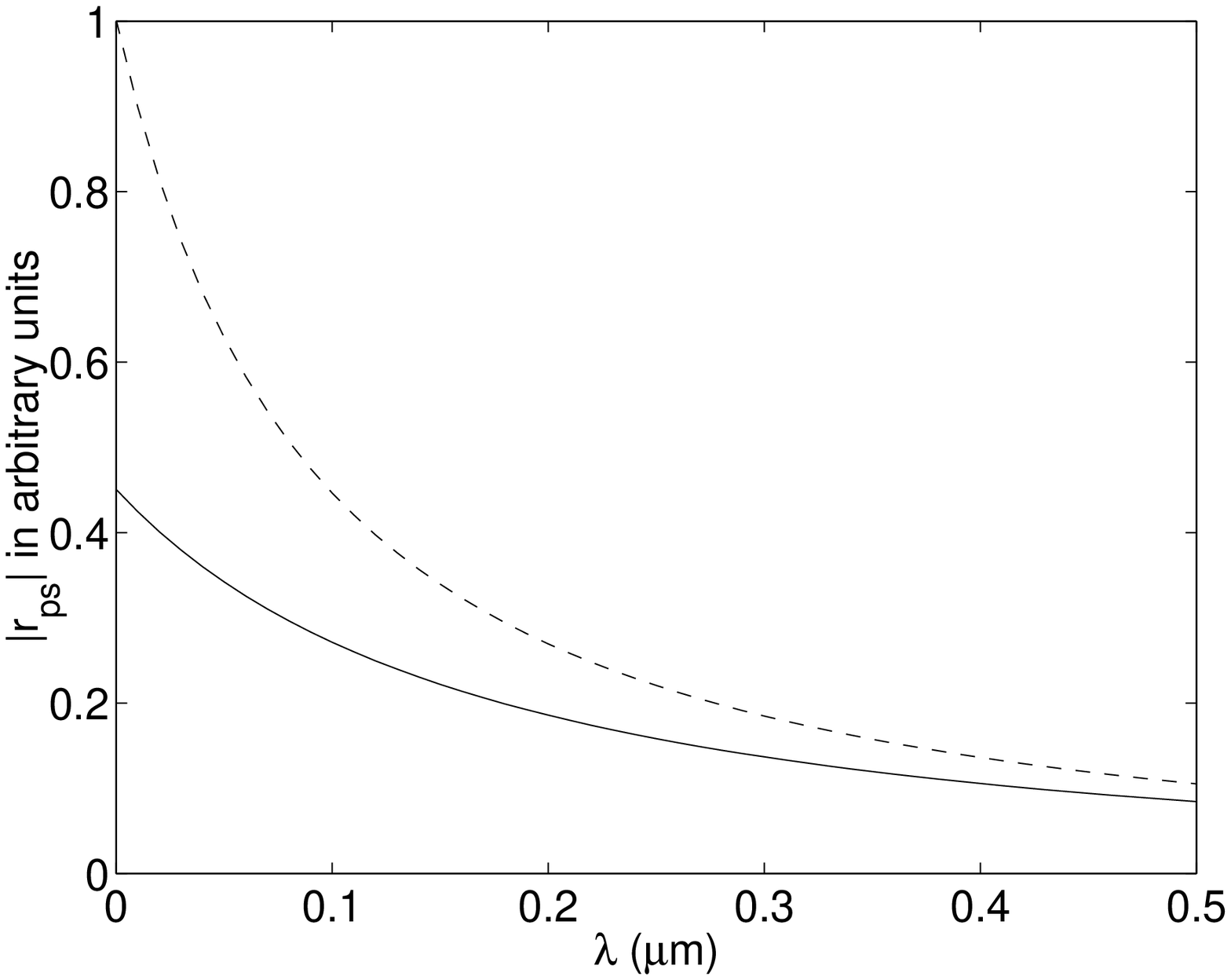}}
\vspace{2cm}
\centerline{Figure~\ref{f3}}

\newpage
\centerline{\includegraphics[width=14cm]{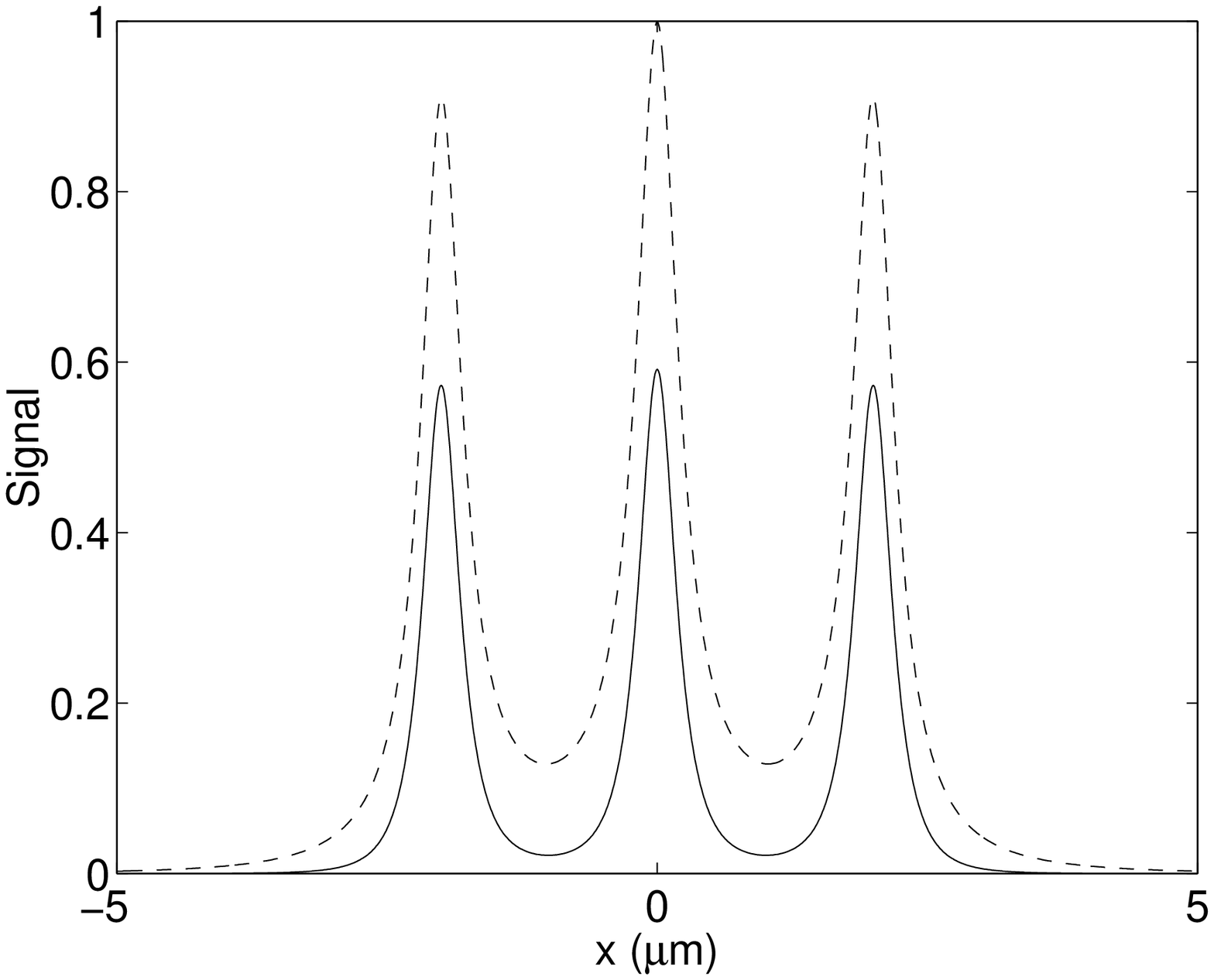}}
\vspace{2cm}
\centerline{Figure~\ref{f4}}


\begin{references}

\bibitem{Alers} P.B. Alers
 {\em Phys. Rev. $\bf{105}$ }, 104 (1957).

\bibitem{Kirchner} H. Kirchner
 {\em Phys. Lett. $\bf{26A}$ }, 651 (1968).

\bibitem{Habermeier1} H.U. Habermeier and H. Kronmuller
 {\em Appl. Phys. $\bf{12}$ }, 297 (1977).

\bibitem{Habermeier2} H.U. Habermeier, W. Klein, R. Aoki and H. Kronmuller
 {\em Phys. Stat. Sol. (a) $\bf{53}$ }, 225 (1979).

\bibitem{Antonov} A.V. Antonov, M.U. Gusev, E.I. Il'yashenko and L.S. Lomov 
 {\em Int. Symp. on Magnetooptics (ISMO'91), USSR, Kharkov} p.70 (1991).

\bibitem{Jevenko} L.A. Jevenko, E.I. Il'yashenko, V.P. Klin, B.P. Nam and A.G. Solovyev 
 {\em Int. Symp. on Magnetooptics (ISMO'91), USSR, Kharkov} p.147 (1991).

\bibitem{Belyaeva} A.I. Belyaeva, A.L. Foshchan and V.P. Yur'ev 
 {\em Sov. Tech. Phys. Lett. $\bf{17}$}, 762 (1991).

\bibitem{Dorosinskii} L.A. Dorosinskii, M.V. Indenbom, V.I. Nikitenko,
Y.A. Ossip'yan, A.A. Polyanskii and V.K. Vlasko-Vlasov
 {\em Physica C $\bf{203}$} (1992), 149 (1992).

\bibitem{Koblischka} M.R. Koblischka and R.J. Wijngaarden ,
 {\em Supercond. Sci. Technol. $\bf{8}$}, 199 (1995).
  
\bibitem{Polyanskii} A.A. Polyanskii, X.Y. Cai, D.M. Feldmann, D.C. Larbalestier,
 {\em NATO Science Series, $\bf{3/72}$}, 353   (Kluwer Academic
 Publishers, Dordrecht 1999).

\bibitem{Runge} B.U. Runge, U. Bolz, J. Eisenmenger and P. Leiderer
 {\em Physica C., $\bf{341}$}, 2029 (2000). 

\bibitem{Goa} P.E. Goa, H. Hauglin, M. Baziljevich, E.I. Il'yashenko, P.L. Gammel
and T.H. Johansen
 {\em Supercond. Sci. Technol., $\bf{14}$}, 729 (2001). 

\bibitem{Helseth} L.E. Helseth, R.W. Hansen, E.I. Il'yashenko, M. Baziljevich and
T.H. Johansen
 {\em Phys. Rev. B, $\bf{64}$}, 174406 (2001).

\bibitem{Vlasko-Vlasov} V.K. Vlasko-Vlasov, Y. Lin, U. Welp, G.W. Crabtree,
D.J. Miller and V.I. Nikitenko
 {\em J. Appl. Phys., $\bf{87}$}, 5828 (2000).

\bibitem{Egorov} A.N. Egorov and S.V. Lebedev 
{\em J. Appl. Phys., $\bf{87}$}, 5362 (2000).

\bibitem{Grechishkin} R.M. Grechishkin, M.Y. Gusev, S.E. Ilyashenko and N.S.
Neustrov 
{\em J. Magn. Magn. Mat. , $\bf{1996}$}, 305 (1996).

\bibitem{Kotov}A. Zvezdin and V. Kotov  
{\em Modern magnetooptics and magnetooptical materials},  (IOP 
Publishing, Bristol, 1997)

\bibitem{Shamonin} M. Shamonin, M. Klank, O. Hagedorn and H. Dotsch 
{\em Appl. Opt., $\bf{40}$}, 3182 (2001).

\bibitem{Novotny} L. Novotny, R.D. Grober and K. Karrai
 {\em Opt. Lett. $\bf{26}$}, 789 (2001). 

\bibitem{Helseth1} L.E. Helseth
 {\em Opt. Commun. $\bf{191}$}, 161 (2001). 

\bibitem{Hubert} A. Hubert and G. Traeger
 {\em J. Magn. Magn. Mater. $\bf{124}$}, 185 (1993). 

\bibitem{Kambersky} V. Kambersky, L. Wenzel and A. Hubert
 {\em J. Magn. Magn. Mater. $\bf{189}$}, 149 (1998).  

\bibitem{Carneiro} G. Carneiro and E.H. Brandt
 {\em Phys. Rev. B 61, 6370 (2000)}. 

\end{references}
\end{document}